\documentclass[fleqn,usenatbib]{mnras}

\usepackage{newtxtext,newtxmath}

\usepackage[T1]{fontenc}
\usepackage{ae,aecompl}

\usepackage{float}

\usepackage{graphicx}   
\usepackage{amsmath}    
\usepackage{amssymb}    
\newcommand{\angstrom}{\mbox{\normalfont\AA}}

\title[Phase-Resolved Spectroscopy of the Low-Mass X-ray Binary V801 Ara]{Phase-Resolved Spectroscopy of the Low-Mass X-ray Binary V801 Ara}

\author[K. Brauer et al.]{
Kaley Brauer,$^{1,2}$\thanks{E-mail: kbrauer@mit.edu}
Saeqa Dil Vrtilek,$^{2}$
Charith Peris$^{2,3}$
and Michael McCollough$^{2}$
\\
$^{1}$Department of Physics \& Kavli Institute for Astrophysics and Space Research, Massachusetts Institute of Technology, Cambridge, MA 02139 USA \\
$^{2}$Harvard-Smithsonian Center for Astrophysics, 60 Garden Street, Cambridge, MA 02138, USA\\
$^{3}$Department of Physics, Northeastern University, 360 Huntington Avenue, Boston, MA 02115, USA
}

\date{Accepted 2018 May 25. Received 2018 May 25; in original form 2017 November 01}

\pubyear{2018}

\begin{document}
\label{firstpage}
\pagerange{\pageref{firstpage}--\pageref{lastpage}}
\maketitle

\begin{abstract}
We present phase-resolved optical spectra of the low mass X-ray binary system V801 Ara. The spectra, obtained in 2014 with IMACS on the Magellan/Baade telescope at Las Campanas Observatory, cover the full binary orbit of 3.8 hours.  They contain strong emission features allowing us to map the emission of H$\alpha$, H$\beta$, He II $\lambda4686$, and the Bowen blend at $\lambda4640$. The radial velocity curves of the Bowen blend shows significantly stronger modulation at the orbital period than H$\alpha$ as expected for the former originating on the secondary with the latter consistent with emission dominated by the disk.  Our tomograms of H$\alpha$ and H$\beta$ are the most detailed studies of these lines for V801 to date and they clearly detect the accretion disk. The H$\beta$ emission extends to higher velocities than H$\alpha$, suggesting emission from closer to the neutron star and differentiating temperature variance in the accretion disk for the first time. The center of the accretion disk appears offset from the center-of-mass of the neutron star as has been seen in several other X-ray binaries. This is often interpreted to imply disk eccentricity.  Our tomograms do not show strong evidence for a hot spot at the point where the accretion stream hits the disk. This could imply a reduced accretion rate or could be due to the spot being drowned out by bright accretion flow around it. There is enhanced emission further along the disk, however, which implies gas stream interaction downstream of the hot spot.
\end{abstract}




\section{Introduction} \label{sec:intro}

V801 Ara (= 4U 1636-536) is a persistent low-mass X-ray binary system (LMXB) comprising a neutron star and a late-type, low-mass ($\sim 0.3-0.4$ $M_{\odot}$) companion star that has filled its Roche lobe and is donating material to the neutron star \citep{1986ApJ...305..246F, 1990A&A...234..181V}.  The system was discovered by \citeauthor{1974RSPSA.340..439W} in 1974 and has a short, well-constrained orbital period of 3.7931206(152) hr \citep{2002ApJ...568..279G}.  Light curve modulations in the system have been attributed to variation from the irradiated companion \citep{1990A&A...234..181V}.  V801 Ara is noted for its frequent burst activity, including ``superbursts'', which are unusually long thermonuclear bursts that can last for hours \citep{1993SSRv...62..223L}.

We observed V801 Ara as part of an ongoing tomographic survey of black hole and neutron star systems in order to search for trends in different classes of X-ray binaries \citep{2015MNRAS.450.2410C,2014ApJ...793...59B,2013ApJ...777..171C,2012MNRAS.427.1043P,2009MNRAS.399..539C, 2007AAS...210.2009N, 2007ApJ...667.1087B, 2004AN....325..209V}. Those studies found that locations of emission and absorption features in the systems depend more strongly on the temperatures of the emitting and absorbing gases and the possibility of disk precession depends more strongly on the mass ratios rather than on the nature of the compact object.

\citet{1998A&A...332..561A} presented radial velocity curves of V801 Ara using the combined fits to He II $\lambda4686$ and the Bowen blend at $\lambda4640$. Doppler tomography of V801 Ara in the emission features He II $\lambda4686$ and the Bowen fluorescence N III $\lambda4640$ were presented by \cite{2006MNRAS.373.1235C}. In addition to revisiting these features we expand on the work of \citet{1998A&A...332..561A} by providing radial velocity curves of fits to the individual lines of H$\alpha$, H$\beta$, He II, and the Bowen blend, and we expand on the work of \cite{2006MNRAS.373.1235C} by presenting the first Doppler tomograms of V801 Ara in H$\alpha$ and H$\beta$.  We also apply modulation tomography \citep{2003MNRAS.344..448S} which allows us to investigate time-dependent variations in all four lines.

In this paper, we present an analysis of optical spectra of V801 Ara obtained in 2014.  Section \ref{sec:obs} describes the observations and data reduction.  Section \ref{sec:analysis} describes the analysis and presents radial velocity curves and tomograms of the stronger features.
Section \ref{sec:disc} compares our tomograms to those of similar systems (persistent binaries containing neutron stars) as well as to transient binaries and systems containing black holes. We summarize our conclusions in Section \ref{sec:conc}.

\section{Observations} \label{sec:obs}

V801 Ara was observed on 2014 May 30 using the Inamori Magellan Areal Camera and Spectrograph (IMACS) on the 6.5m Walter Baade Telescope at Las Campanas observatory.  A 600 mm\textsuperscript{-1} grating was used, providing a dispersion of 0.378 $\angstrom$ per pixel covering a wavelength range from 4450-6700 {\AA} (with small gaps due to chip edges).  A 1 arcsec slit was used throughout, resulting in a spectral resolution of 3.4 {\AA} and a velocity resolution of 175 km s\textsuperscript{-1}. The seeing ranged from 0.6-1.25 arcsec during our observations.  A total of 39 spectra were obtained.  We applied no binning, but used a four chip sub-array to reduce readout time.  The spectra had integration times of 300-600s and covered 1.8 binary orbits (Table \ref{tab:obs}).  He+Ne+Ar comparison lamps were taken for every 3-4 exposures on target.  A 3rd-order fit to a total of 21 lines across the spectrum gave us rms scatter less than 0.005 \AA. 

The spectra were reduced with the IRAF IMACS package \footnote{IRAF is distributed by the National Optical Astronomy Observatory, which is operated by the Association of Universities for Research in Astronomy (AURA) under cooperative agreement with the National Science Foundation.}.  The reduction included applying bias subtraction and flat fielding to the multi-dimensional frames, extracting the one-dimensional spectra, and conducting wavelength calibration on the extracted spectra using comparison arcs nearest in time to the observation.

\section{Analysis and Results} \label{sec:analysis}

\subsection{Spectral Features} \label{subsec:specV801}

\begin{figure*}
\centering
    \includegraphics[width=0.9\textwidth]{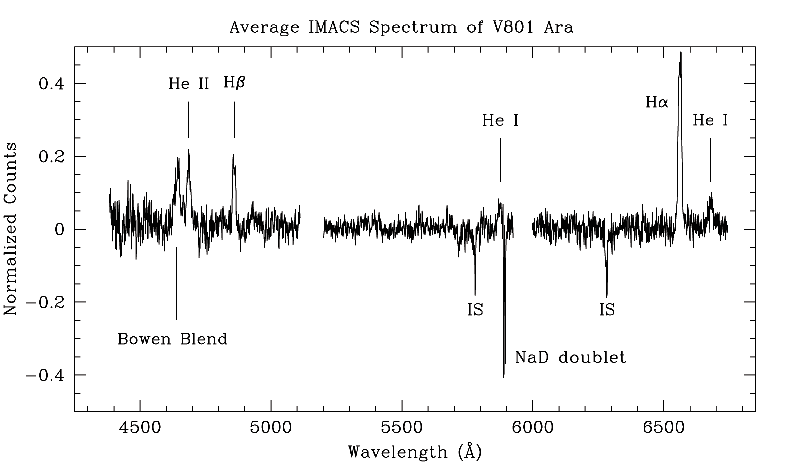}
    \caption{Continuum-subtracted spectrum that is the mean of 39 individual spectra recorded on 2014 May 30. The major lines, including H$\alpha$, H$\beta$, He II $\lambda 4686$, He I $\lambda 5876$, He I $\lambda 6678$, and the Bowen blend at $\lambda\lambda 4634-4651$ (a blend of N III $\lambda\lambda4634-41$ and C III $\lambda\lambda4647-51$), are identified. Blank spaces in the spectrum are due to chip gaps. \label{fig:V801AvSpec}}
\end{figure*}

The spectra of V801 Ara display strong emission features of H$\alpha$, H$\beta$, He II $\lambda 4686$ and the Bowen blend at $\lambda 4634-4651$ (primarily a blend of N III $\lambda\lambda4634-41$ and C III $\lambda\lambda4647-51$) and weaker features of He I $\lambda 5876$ and He I $\lambda6678$.  These lines are typical of X-ray active LMXBs and are consistent with lines that were identified by \citet{2006MNRAS.373.1235C} and \cite{1998A&A...332..561A}.  Figure \ref{fig:V801AvSpec} shows the mean continuum-subtracted spectrum with the major lines identified.

\begin{figure*}
\centering
    \includegraphics[width=0.95\textwidth]{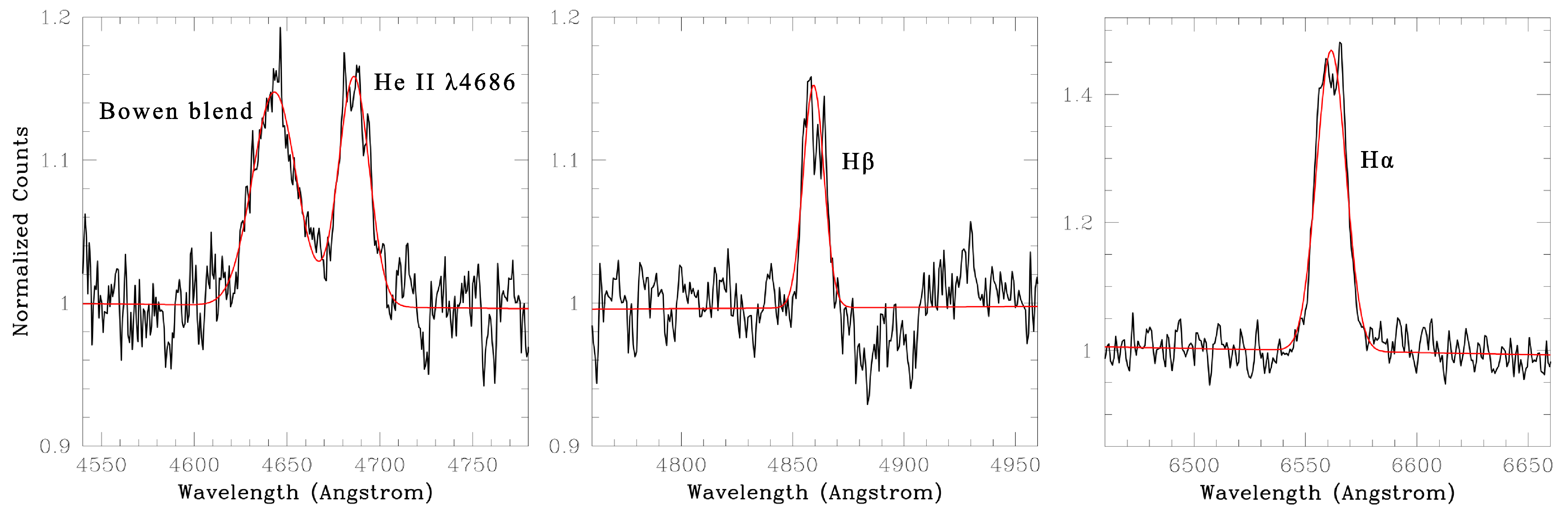}
    \caption{Each emission feature chosen for tomographic analysis is shown with a Gaussian fit in red overlaying the normalized mean spectrum. Each data point represents a pixel with 0.378 {\AA}  resolution. \label{fig:V801gaus}}
\end{figure*}

The H$\alpha$, H$\beta$, He II $\lambda 4686$, and Bowen blend were each fit by a Gaussian.  Details of the fits are displayed in Figure \ref{fig:V801gaus}, and the associated line parameters are listed in Table \ref{tab:V801Line}. H$\alpha$, H$\beta$ and He II $\lambda 4686$ display clear double peaks which are characteristic of emission from a disk. Our measured equivalent widths (EWs) and full widths at half maximum (FWHMs) for H$\beta$, He II, and Bowen blend agree with those measured by \citet{2006MNRAS.373.1235C} with the exception that our H$\beta$ FWHM is smaller. The EWs are significantly larger than those measured by \cite{1998A&A...332..561A}, suggesting variability in the system on long timescales. Because the EWs do not significantly vary as a function of phase, short timescale variability would not account for the differences. \cite{1998A&A...332..561A} did not report FWHMs or centroids, and neither \cite{1998A&A...332..561A} nor \citet{2006MNRAS.373.1235C} observed H$\alpha$.

The Rossi X-Ray Timing Explorer (RXTE) All-Sky Monitor (ASM) light curve of V801 Ara over a period of 15 years can be seen in Figure \ref{fig:rxte}. The X-ray counts are highly variable, consistent with variability in the system. Optical light from X-ray binaries is strongly influenced by X-ray heating of the accretion disk and surface of the neutron star \citep{1990A&A...235..162V,1993ApJ...404..696V}. The factor of four changes in X-ray flux for V801 can easily explain the differences in optical line strengths.

\begin{table}
    \caption{Emission Line Parameters from Gaussian Fits \label{tab:V801Line}}
    \begin{tabular}{cccc}
    \hline
    Line & FWHM & $\textrm{EW}$ & Centroid \\
     & (\AA) & (\AA) & (\AA) \\
    \hline
    Bowen blend & 27.83 $\pm$ 0.57 & 4.27 $\pm$ 0.08 & 4642.30 \\
    He II $\lambda 4686$ & 19.11 $\pm$ 0.69 & 3.23 $\pm$ 0.05 & 4685.79 \\
    H$\beta$ & 10.92 $\pm$ 0.23 & 1.71 $\pm$ 0.05 & 4859.71 \\
    H$\alpha$ & 15.55 $\pm$ 0.09 & 7.62 $\pm$ 0.10 & 6561.77 \\
    \end{tabular}
\end{table}

\subsection{Radial Velocities} \label{subsec:RadVel}

\begin{figure*}
\centering
    \includegraphics[width=0.7\textwidth]{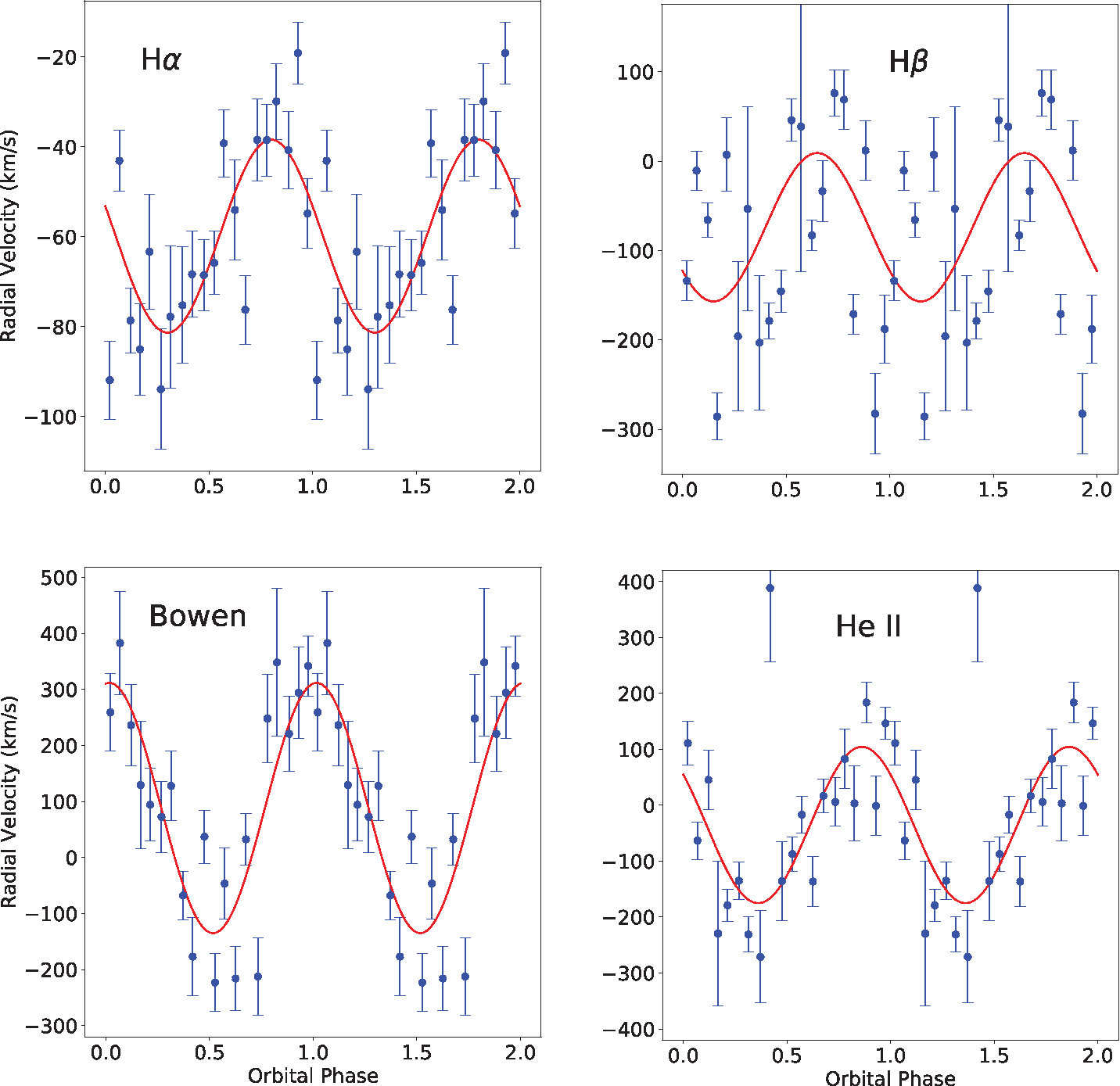}
    \caption{Binned radial velocity measurements versus orbital phase. The data are overlayed with the fits used to obtain $K$, $\gamma$, and $\phi_0$ values (with $V(\phi) = \gamma + K \sin[2\pi (\phi - \phi_0)]$).  The orbital cycle is repeated for clarity. \label{fig:radvel}}
\end{figure*}

Radial velocities of the lines listed in Table \ref{tab:V801Line} were computed by cross-correlating individual spectra with Gaussians at the listed FWHM and line centroids and are displayed in Figure \ref{fig:radvel}. The velocities are binned and folded using phases determined by using the spectrophotometric ephemeris from \citet{2006MNRAS.373.1235C}: 

\begin{equation}
T_0 (HJD) = 2452813.531(2) + 0.15804693(16) E
\end{equation}

where $T_0$ is defined as the compact object superior conjunction and 0.15804693(16) is the orbital period in days.  The uncertainty in phase for the time of our observations amounts to $\pm$0.02 orbital cycles. We note that \citet{2006MNRAS.373.1235C} set the zero point of their ephemeris by requiring that a hot spot observed in their Bowen line emanated from the heated face of the secondary.

We fit each radial velocity curve with $V(\phi) = \gamma + K \sin[2\pi (\phi - \phi_0)]$ to determine $K$, $\gamma$, and $\phi_0$ values.  The fits are shown in Figure \ref{fig:radvel} and the corresponding values can be found in Table \ref{tab:radvel}.

The radial velocity curves of H$\alpha$ and He II show lower excursions than that of the Bowen blend consistent with associating them with emission from the disk around the more massive compact object (Casares et al. found that $K_1 / K_2 = M_2 / M_1 < 1$, so the compact object is more massive than the companion star). The H$\beta$ radial velocity curve is contaminated by strong phase dependent absorption features longward of the line making it difficult to detect the underlying orbital modulation of the system.  Because of this, the fit shown in red for this line is poor.  The radial velocities of the Bowen blend cannot be easily interpreted as they come from a complex blend of emission features, but we present them here in comparison with the radial velocities measured by \citet{2006MNRAS.373.1235C}. Both He II and the Bowen complex show the same velocity excursions observed by Casares et al. for V801 Ara, and the phase of maximum velocity for the Bowen complex (at phase 1.0) also agree.  The phases of maximum radial velocities for H$\alpha$, H$\beta$, and He II $\lambda4686$, are offset from the phase of the maximum of the Bowen blend as one would expect if the former represent emission from the disk (consistent with their double-peaked structure) and the latter represents emission from the secondary.


\begin{table}
    \caption{Radial Velocity Single Gaussian Fits \label{tab:radvel}}
    \begin{tabular}{cccc}
    \hline
    Line & $\gamma$ & $K$ & $\phi_0$  \\
     & (km s\textsuperscript{-1}) & (km s\textsuperscript{-1}) &  \\
    \hline
     Bowen blend & 88.5 $\pm$ 25.0 & 223.3 $\pm$ 34.4 & 0.77 $\pm$ 0.03 \\
    He II $\lambda 4686$ & -35.6 $\pm$ 23.6 & 139.8 $\pm$ 34.6 & 0.61 $\pm$ 0.04 \\
    H$\beta$ & -90.2 $\pm$ 24.4 & 62.3 $\pm$ 34.5 & 0.44 $\pm$ 0.08 \\
    H$\alpha$ & -59.8 $\pm$ 3.8 & 21.5 $\pm$ 5.7 & 0.55 $\pm$ 0.04 \\
    \end{tabular}
\end{table}

\begin{figure*}
\centering
\includegraphics[width=0.75\textwidth]{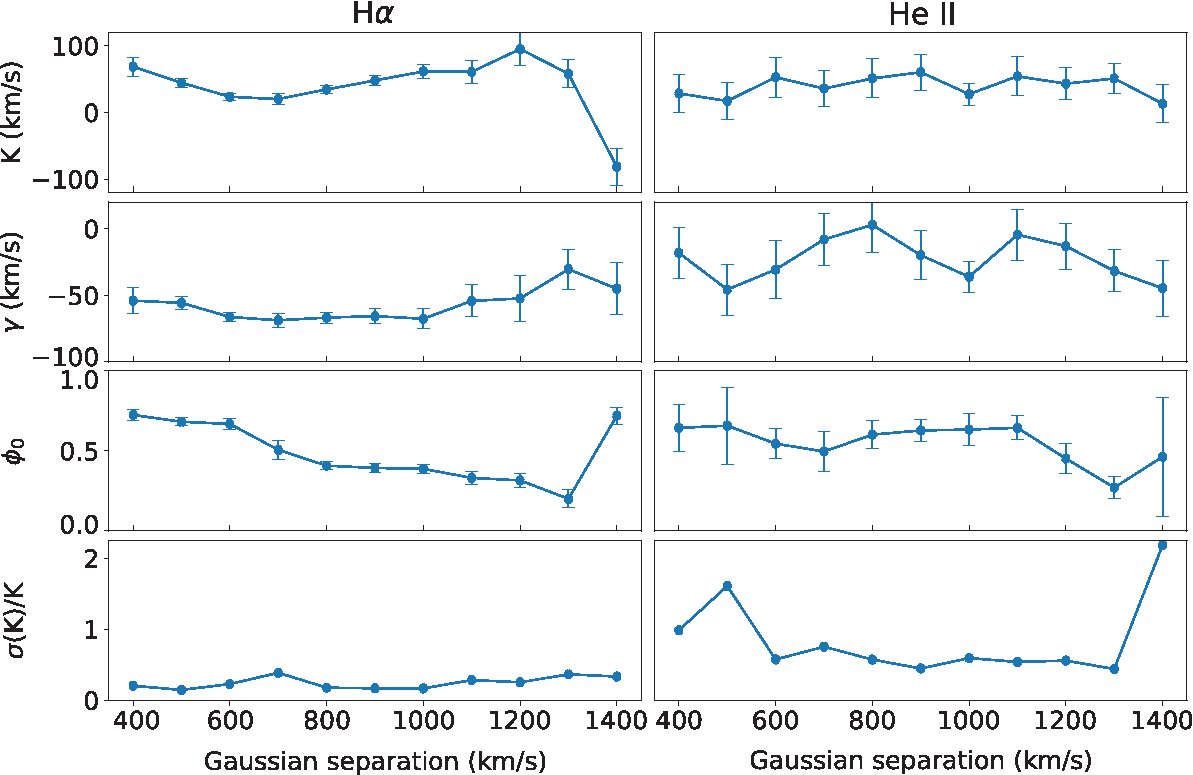}
\caption{Diagnostic diagrams for H$\alpha$ and He II $\lambda 4686$ obtained by applying the double-Gaussian technique with separations in steps of 100 km s\textsuperscript{-1}. Error bars do not include possible systematics and thus may underestimate the actual error. \label{fig:rvelwing}}
\end{figure*}

To supersede the inconsistent parameters from our core analysis, we have also applied the double-Gaussian technique \citep{1980ApJ...238..946S} to the wings of the H$\alpha$ and He II lines.  This technique lets us extract radial velocity curves from the wings of the line profile by convolving the emission with double-Gaussian filters of varying Gaussian separation.  Because the wings of these lines are expected to follow the motion of the neutron star, extracting these radial velocity curves allows us to estimate system parameters.  For both lines, we used Gaussians with FWHM = 100 km s\textsuperscript{-1} and relative Gaussian separations of $400 - 1400$ km s\textsuperscript{-1} in steps of 100 km s\textsuperscript{-1}.  The radial velocity curves for each Gaussian separation were fit with $V(\phi) = \gamma + K \sin[2\pi (\phi - \phi_0)]$.  Diagnostic diagrams showing the corresponding best-fit parameters are presented in Figure \ref{fig:rvelwing}.

\begin{table}
    \caption{Radial Velocity Double-Gaussian Fits of H$\alpha$ \label{tab:doubleG}}
    \begin{tabular}{cccc}
    \hline
    Separation & $\gamma$ & $K$ & $\phi_0$  \\
     (km s\textsuperscript{-1}) & (km s\textsuperscript{-1}) & (km s\textsuperscript{-1}) &  \\
    \hline
     700 & -68.8 $\pm$ 5.3 & 20.0 $\pm$ 7.7 & 0.51 $\pm$ 0.06 \\
     800 & -66.7 $\pm$ 4.1 & 35.4 $\pm$ 6.2 & 0.41 $\pm$ 0.02 \\
    900 & -65.6 $\pm$ 5.7 & 47.9 $\pm$ 8.1 & 0.39 $\pm$ 0.03 \\
    1000 & -67.6 $\pm$ 7.4 & 61.5 $\pm$ 10.4 & 0.39 $\pm$ 0.04 \\
    1100 & -54.2 $\pm$ 12.0 & 60.8 $\pm$ 17.4 & 0.34 $\pm$ 0.03 \\
    \end{tabular}
\end{table}

Our parameters for He II do not agree with those found by \citet{2006MNRAS.373.1235C} using the same technique. As they note, however, the greatly asymmetric distribution of the He II Doppler tomogram (see next section) likely invalidates the double-Gaussian technique for this line.  The H$\alpha$ tomogram is more symmetric, and the $\gamma$ value obtained from the double-Gaussian wing analysis agrees with the value obtained from the H$\alpha$ core analysis.  For both these reasons and because the H$\alpha$ line is isolated and much stronger than He II, we used H$\alpha$ to obtain parameter estimates. We used data points at Gaussian separations of $800 - 1000$ km s\textsuperscript{-1} because they were associated with the lowest errors (in the diagnostic diagram, the bottom of Figure \ref{fig:rvelwing}) and displayed the most stable values. The data points for these separations and the surrounding separations can be found in Figure \ref{tab:doubleG}.

We obtained a radial velocity amplitude of the neutron star $K_1 = 26 $ to $ 57$ km s\textsuperscript{-1}, a systemic velocity $\gamma = -79 $ to $ - 57$ km s\textsuperscript{-1}, and a zero-phase $\phi_0 = 0.35$ to $0.46$. The ranges given are 3 sigma ranges since the standard deviation errors do not include possible systematic errors and thus may underestimate the true uncertainty. While we used the range of separations with the lowest errors and greatest stability, even across this range the K1 values display instability.  For these reasons and because of the moderate spectral resolution and slight asymmetry in the tomogram, these values should be taken with caution.


\subsection{Tomography} \label{subsec:dopTom}

Doppler tomography \citep{2016..ASSL..439M, 2005Ap&SS.296..403M} can be used to probe the structure and variability of line-emitting or line-absorbing regions in binary systems.  Tomograms are two-dimensional maps in velocity space requiring line profiles obtained at frequent intervals around a full orbital period.  The Doppler broadening of the line profile contains information about the velocity distribution, and the rotation of the system supplies a continuous series of projections from which a tomogram can be formed.

Standard Doppler tomography produces time-averaged tomograms.  Modulation Doppler tomography \citep{2003MNRAS.344..448S}, an extension of standard Doppler tomography, allows for investigation of time-dependent variations in the line profiles as well.  \cite{2003MNRAS.344..448S} used an iterative reduced $\chi^2$ process that modified a simple starting image (a uniform grid or a Gaussian) to fit the data by gradually decreasing the reduced $\chi^2$ in comparison to the data.  Because many possible fits exist for a given $\chi^2$, the Maximum Entropy Method \citep{1986ARA&A..24..127N} was used to select the most realistic fit.  The Maximum Entropy Method, based on the idea that higher entropy corresponds to a smoother and more physically probable image, involves selecting the fit with the highest entropy after each iteration. Modulation tomography takes into account how the flux from a given point varies as a function of time and produces a tomogram that takes this into consideration showing emission sources that harmonically vary as a function of the period.  

We used Tom Marsh's MOLLY package to prepare spectra for  tomography\footnote{deneb.astro.warwick.ac.uk/phsaap/software/molly/html/INDEX.html}. Heliocentric corrections were made to take into consideration the location of the telescope on Earth and Earth's velocity. We fitted curves to the continuum while excluding strong features and subtracted the fit from each spectrum.  We then isolated each emission feature we planned to analyze by masking all portions of the spectra except for the relevant line and some continuum around it.

\begin{figure*}
\centering
    \includegraphics[width=0.9\textwidth]{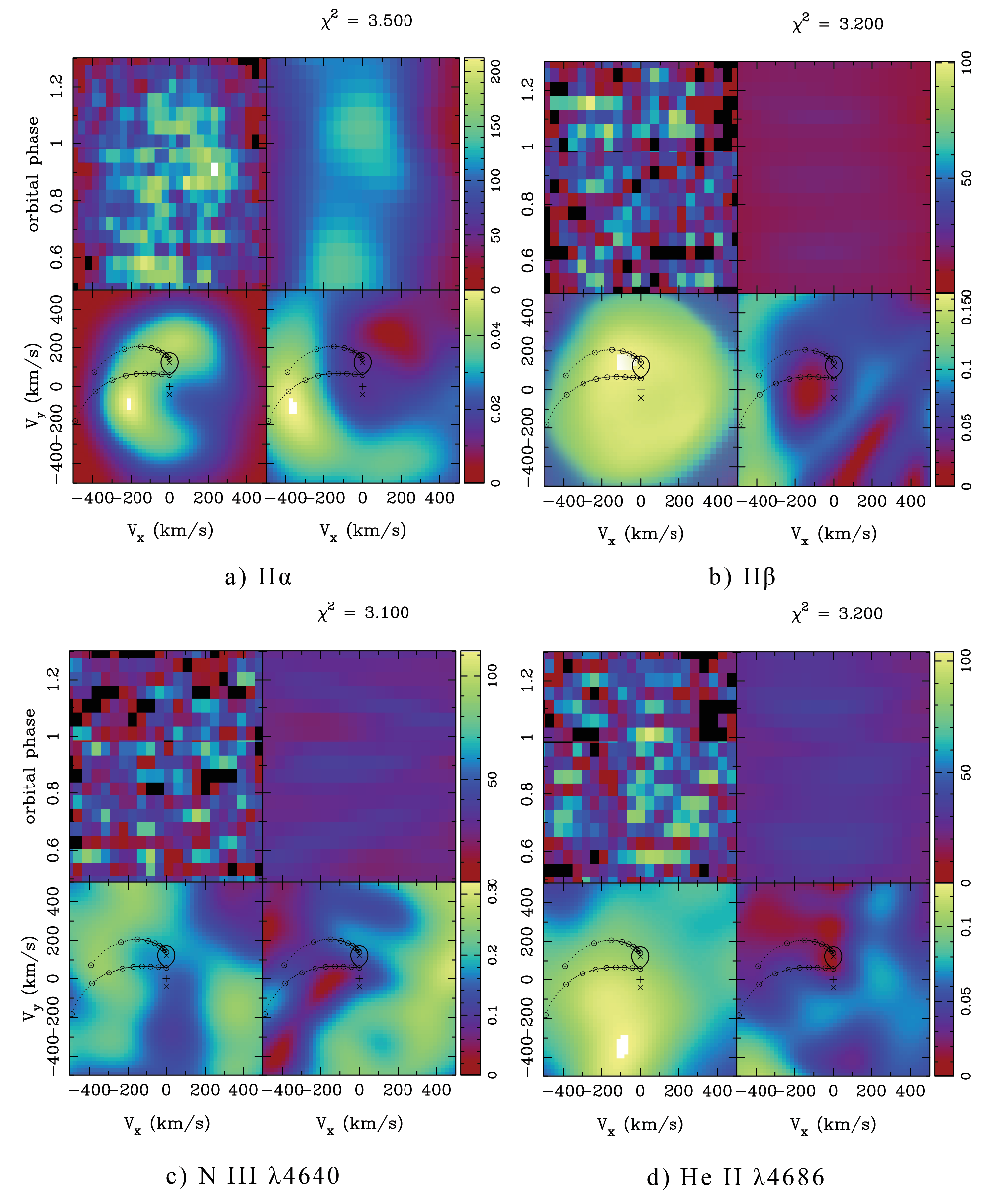}
    \caption{Modulation tomograms of the H$\alpha$, H$\beta$, N III $\lambda4640$, and He II $\lambda4686$ lines in V801 Ara. These maps assume $\gamma = -68$  km s\textsuperscript{-1} from Section \ref{subsec:RadVel}. They also use $K_1 = 42$ km s\textsuperscript{-1} from Section \ref{subsec:RadVel} along with an assumed a mass ratio of 0.27 to plot expected locations of the components of the system. In each set of four panels, the top left shows the trailed observed data and the top right shows a trail that was recreated using the tomogram. The bottom left shows a map of the constant emission of the accretion disk and the bottom right shows a map of the modulation amplitude. For the trailed spectra and the constant emission tomogram, lighter color indicates higher intensity. For the modulation tomogram, lighter colors indicate a greater modulation amplitude. The companion star's Roche lobe (a teardrop shape) and the predicted centers of mass of the system (+), neutron star (lower x), and the companion star (upper x) are plotted on the tomograms. The lower curved line represents the ballistic trajectories of the accretion stream, and the upper curved line represents the Keplerian velocity of the disk along the stream. Each open circle along the curves demarks $0.1 R_{L1}$ where $R_{L1}$ is the distance from the compact object to the inner Lagrangian point.\label{fig:V801tom}}
\end{figure*}

\begin{figure*}
\centering
    \includegraphics[width=0.9\textwidth]{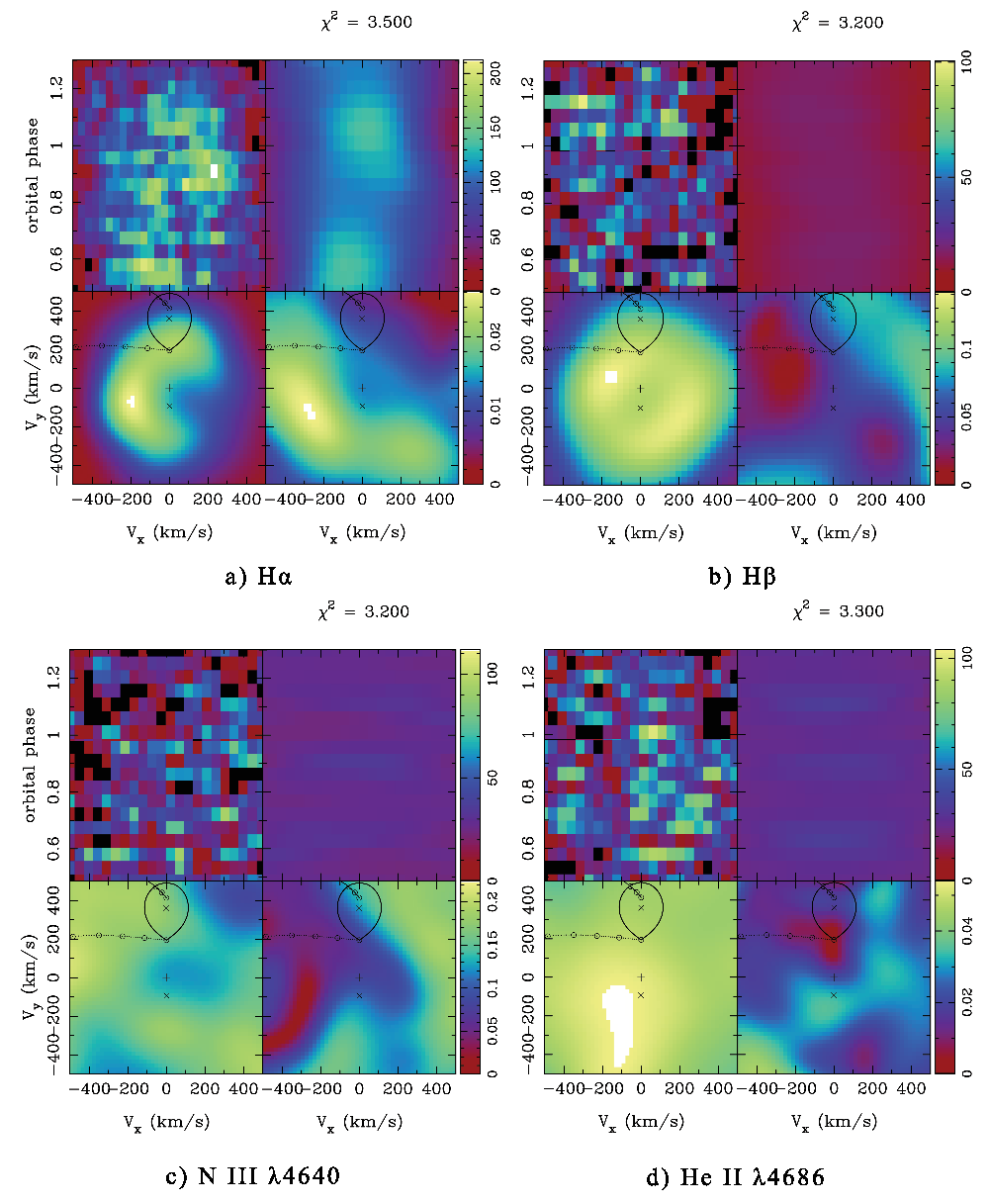}
        \caption{As in Fig. \ref{fig:V801tom} but with assumed system parameters ($\gamma = -34$ km s\textsuperscript{-1}, $K_1 = 90-113$ km s\textsuperscript{-1}, $K_2 = 360$ km s\textsuperscript{-1}) used by \citet{2006MNRAS.373.1235C}. \label{fig:V801tomCas}}
\end{figure*}

Figure \ref{fig:V801tom} shows tomograms and phase-folded trailed spectra of V801 Ara that were computed for the emission features H$\alpha$, H$\beta$, He II $\lambda 4686$, and the Bowen blend at $\lambda\lambda 4634-4651$.  To create the maps, we assumed a systemic velocity of $\gamma = -68$  km s\textsuperscript{-1} as measured in Section \ref{subsec:RadVel}.  On each time-averaged tomogram, we plotted the companion star's Roche lobe and the centers of mass of the neutron star (x), companion star (x), and the system (+). These were predicted based on $K_1 = 42$ km s\textsuperscript{-1} from the range in the previous section and a mass ratio of $q = 0.27$ from the range given by \citet{2006MNRAS.373.1235C}.  The ballistic trajectories of the accretion stream and the Keplerian velocity of the disk along the stream are also both plotted over each tomogram of constant emission as the lower and upper curved lines, respectively.

The Doppler tomograms of constant emission (the lower left panel in each plot) of H$\alpha$ and H$\beta$ (Figure \ref{fig:V801tom}a and b) clearly show a ring of high intensity that corresponds to the accretion disk.  For H$\alpha$, which appears to originate from material close to the outer edge of the disk, the disk appears to be slightly offset from the center-of-mass of the neutron stars towards the center-of-mass of the system and is somewhat asymmetric, whereas H$\beta$ shows higher velocities (associated with material closer to the center of the disk) and a more symmetric disk. H$\alpha$ does not show a hot spot where we expect the accretion stream to be hitting the disk (this should appear between the ballistic trajectories of the accretion stream and the Keplerian velocity of the disk along the stream which are plotted as curved lines on the tomograms), but H$\beta$ shows a slight hot spot.  H$\alpha$, our stronger and more reliable tomogram, shows enhanced emission below the expected position of the hot spot.


As mentioned earlier, the H$\beta$ line is contaminated by phase-variable absorption approximately $30 \angstrom$ redward of the centroid.  Phase-dependence violates the principles of standard Doppler tomography, but modulation tomography accounts for flux that varies with time.  We thus used modulation Doppler tomography to produce tomograms of constant emission and varying emission for each emission feature. Still, modulation Doppler tomography is less reliable when studying complex profiles comprised of emission and variable absorption, so our $H\beta$ tomogram should be treated with caution.

The tomograms of the Bowen fluorescence using $\lambda4641.66$ as the centroid and of He II $\lambda4686$ are shown in Figure \ref{fig:V801tom}c and d. The tomogram of He II $\lambda4686$ shows no hot spot, no emission from the secondary, and emission from the outermost parts of the disk (ie., material at lower velocities).  The Bowen blend tomogram (centered on $\lambda 4641.66$) shows evidence of emission from the companion star, but there is additional emission from regions not associated with the companion.

In Figure \ref{fig:V801tomCas} we show tomograms made using the system parameters obtained by \citet{2006MNRAS.373.1235C}. These tomograms assume a systemic velocity of $\gamma = -34 \pm 5$  km s\textsuperscript{-1}, a $K_2$ value of $360 \pm 74$ km s\textsuperscript{-1}, and an average of the range ($90-113$ km s\textsuperscript{-1}) they provided for $K_1$. We tested the full range of $K_1$ and $K_2$ values that were given by Casares et al. and found differences to be less than 10 percent.

The tomographic images created with the system parameters of \citet{2006MNRAS.373.1235C} are nearly identical to those created with our newly obtained system parameters, with only a few important differences. These differences are primarily due to the different values of $K_1$ and $K_2$; the different $\gamma$ values produce holistically similar tomograms.

The system parameters of Casares et al. result in disks which appear to be centered on the center-of-mass of the system, particularly in the H$\alpha$ tomogram.  This H$\beta$ tomogram shows slightly enhanced emission below the expected position of the hot spot, as also seen in both H$\alpha$ tomograms. Notably, the disk components seen in the H$\alpha$ and H$\beta$ tomograms have velocities lower than the $K_2$ value.  As we discuss in the next section, this could imply that these $K_1$ and $K_2$ values are overestimates.


\section{Discussion} \label{sec:disc}

Here we discuss possible implications of the features found in our V801 Ara tomograms. These are discussed in context with tomograms of other LMXBs containing both neutron stars and black holes. We also describe what other observations would be necessary to confirm different interpretations.

The H$\alpha$ tomogram in Figure \ref{fig:V801tom} shows a disk that appears to be centered on the center-of-mass of the system slightly more than that of the neutron star. This small offset could imply disk eccentricity. (Note: the larger offset seen in Figure \ref{fig:V801tomCas} using values from \citet{2006MNRAS.373.1235C} cannot be completely explained by an eccentric disk.) Tomograms of systems both similar to V801 Ara (e.g. X1822-371 \citet{1997MNRAS.285..673, 2012MNRAS.427.1043P}), and other systems in quiescence containing both neutron stars (e.g. Cen X-4, \citet{2005A&A.444.905}) and black holes (e.g. A0620, \citet{2008MNRAS.384.849N}) are also consistent with an asymmetric disk. \citet{2012MNRAS.427.1043P} suggested that the combination of a disk offset from the predicted position of the neutron star and the asymmetric structure could be the indication of an eccentric disk.  An eccentric disk implies disk precession. In the case of A0620-00, \citet{2008MNRAS.384.849N} propose that the precession is due to a gravitational torque from the companion star that introduced asymmetries and shifts of the disk.  Because the system parameters of V801 Ara ($K_1$, $K_2$, and $\gamma$ velocity) are not well determined, however, we cannot say with any certainty that we have evidence for an eccentric disk.   Deviations in the centroids of double peaked lines can also be used to determine disk eccentricity.  Since this is the first measurement of the line centroid of H$\alpha$ for this object and there exist only two measurements of the He II line centroid (\citet{2006MNRAS.373.1235C} and this paper) which span 25 years, we would need more data before we could either support or reject a precessing disk.

Our tomograms (H$\alpha$, H$\beta$, He II, and Bowen) do not show evidence for a hot spot, except for a slight enhancement in the H$\beta$ tomogram with our system parameters.  The lack of a hot spot associated with the impact point of the accretion stream on the disk in Doppler tomograms and excess emission below the expected position of the hot spot has been observed in several systems: V926 Sco \citep{2013ApJ...777..171C}; EXO0748-676 \citep{2012ApJ...750..132M, 2009MNRAS.394L.136M}; and Cen X-4 \citep{2005A&A.444.905,2003MNRAS.341.1231T}.  Several interpretations exist for the lack of a hot spot.  Torres et al. suggest that the absence of a hot spot in H$\alpha$ is due to Cen X-4 underfilling its Roche lobe making mass transfer unstable.  D'Avanzo et al. saw no hot spot with H$\alpha$ but did see it with He I $\lambda5876$ and suggest that in Cen X-4 the temperature at the hot spot is too high to produce H$\alpha$.  
The lack of a hot spot in V926 Sco was interpreted by \citet{2013ApJ...777..171C} as an indication of a reduced accretion rate; this was consistent with simultaneous X-ray observations with RXTE that show reduced X-ray emission during the optical observations (in contrast with earlier optical observations that showed the hot spot).  Both Torres and D'Avanzo interpretations are consistent with the variable mass transfer inferred by Connolly et al.  Previous X-ray observations (Figure \ref{fig:rxte}) show that V801 Ara has a varying accretion rate, so the possibility of it being in a state of reduced accretion is reasonable.  On the other hand, if the system is in a high accretion state, the lack of a clear hot spot could possibly be due to the spot being drowned out by the bright accretion flow around it. We would need simultaneous X-ray observations of V801 Ara to confirm or deny these interpretations.  As for the enhancement observed below the expected position of the hot spot, this has repeatedly been interpreted as due to the gas stream re-impacting the disk \citep{2013ApJ...777..171C, 2012ApJ...750..132M, 2009MNRAS.394L.136M, 2002MNRAS.329...29C}.

He II $\lambda 4686$ and Bowen line tomograms have been reported for several systems similar to V801 Ara.  For these systems, the majority of tomograms of He II $\lambda4686$ are consistent with disk emission and only a few cases also show emission from the secondary.  Bowen blend emission is generally associated with the secondary.  These studies thus support our association of He II with the disk and Bowen blend from the secondary.  Many of these tomograms also exhibit similar features to ours.  Emission from the He II $\lambda4686$ maps of Sco X-1 \citep{2002ApJ...568..273S} and of X1822-371 \citep{2003ApJ...590...1041C} show no hot spot, excess emission below the hot spot, and are consistent with an asymmetric disk.  However, \cite{2002ApJ...568..273S} also show strong He II emission from the secondary although they note in their text that their trailed spectra of He II $\lambda4686$ is anti-phased with the Bowen blend components and therefore must be associated with the compact object. \cite{2003ApJ...590...1041C} show Bowen emission only from the secondary.  
The He II $\lambda4684$ map of EXO 0748-076 by \cite{2009MNRAS.394L.136M} is similar to ours in showing no hot spot and excess emission below the hot spot, but they also show enhanced emission from the secondary.  The \citet{2012ApJ...750..132M} tomogram of He II $\lambda4686$ for EXO 0748-076 shows no hot spot and asymmetric disk emission as we see, but also a suggestion of emission from the heated face of the secondary.  The He II $\lambda 4686$ tomograms of AC211 \citep{2003MNRAS.341.1231T} show both enhanced emission from the secondary and a faint disk centered on the center-of-mass of the system.

When reporting their tomograms of V801 Ara, \citet{2006MNRAS.373.1235C} also reported tomograms of V926 Sco in He II $\lambda 4686$ and N III $\lambda 4640$. We note that these lines were revisited by \cite{2013ApJ...777..171C} using the IMACS, and while their He II tomogram matched that of \citet{2006MNRAS.373.1235C}, their N III tomogram did not. The tomograms of both these lines in X1822-371 by \cite{2012MNRAS.427.1043P} (again using IMACS) and \cite{2003ApJ...590...1041C} are well matched.  Our and \cite{2013ApJ...777..171C}'s lack of clear Bowen emission from the companion star, in contrast to that reported by Casares et al., is likely due to a combination of the weakness of these sources relative to X1822-371 as well as our lower spectral resolution.  For the weaker sources, IMACS requires integration times almost twice as long as for X1822-371. This smears the orbital phase resolution.  The lower spectral resolution makes it difficult to assess the presence of significant emission for the narrow, weak Bowen lines.  

As mentioned in the previous section, our tomograms may suggest that the $K_1$ and $K_2$ values used by \citet{2006MNRAS.373.1235C} are overestimates. The disk components seen in the H$\alpha$ and H$\beta$ tomograms using the system parameters from Casares et al. have velocities lower than the $K_2$ value. Kepler's law tells us that the disk should be moving faster than $K_2$, so emission components that lie at projected velocities lower than $K_2$ cannot be easily attributed to the disk without involving highly sub-Keplerian flows.  Because we expect this emission to be coming from the disk, this implies that the $K_1$ and $K_2$ values obtained in Casares et al. are an overestimate.  The $K_2$ implied by our $K_1$ value and a mass ratio of 0.27 is lower than that of Casares et al. and does not conflict with the H$\alpha$ and H$\beta$ emission being attributed to the disk.

There is still much uncertainty regarding V801 Ara's $K_1$ and $K_2$ values, however. The higher values reported in Casares et al. could still be explained by highly sub-Keplerian or non-disk flows. Because $K_1$ and $K_2$ are so important in identifying the location of emission in the maps, more work needs to be done to narrow down these values.

\section{Conclusions} \label{sec:conc}

The optical spectra of V801 Ara obtained with IMACS display strong emission features at H$\alpha$, H$\beta$, He II $\lambda4686$, and the Bowen fluorescence at $\lambda4634-4651$.  H$\alpha$, H$\beta$, and He II $\lambda4686$ clearly show double-peaked structure consistent with emission from a disk.

This is the first determination of the radial velocities of H$\alpha$ and we find that they show considerably less excursion than that of the Bowen blend. This is consistent with H$\alpha$ emission dominantly from the disk (hence associated with the more massive neutron star) and Bowen emission associated with the less massive secondary.  The phases of maximum radial velocities for H$\alpha$, H$\beta$, and He II $\lambda4686$, are offset from the phase of the maximum of the Bowen blend, consistent with the former representing emission from the disk and the latter representing emission from the secondary. The radial velocity curves we determined for He II $\lambda 4686$ and the Bowen blend are consistent with those determined by \citet{2006MNRAS.373.1235C}.  We also obtain new estimated ranges for $K_1 = 26$ to $57$ km s\textsuperscript{-1} and $\gamma = -79$ to $-57$ km s\textsuperscript{-1} using the double-Gaussian method on the H$\alpha$ line. This range of $K_1$ values, which is lower than that found by \citet{2006MNRAS.373.1235C}, better agrees with the expected attribution of H$\alpha$ and $H\beta$ emission to the neutron star's accretion disk. These estimates for the system parameters should be treated with caution, however, due to possible systematic errors, our moderate spectral resolution, and the slight asymmetry in the H$\alpha$ tomogram. There is still much uncertainty in the $K_1$ and $K_2$ values.

The tomogram of N~III $\lambda 4640$ (from the Bowen blend) shows some emission that can be associated with the secondary but also shows emission from regions not associated with the secondary. This is likely due to our lower spectral resolution not well capturing the narrow, weak Bowen emission. None of the tomograms show excess emission at the expected position of the spot where the accretion stream hits the disk except a slight enhancement in the H$\beta$ tomogram created with our system parameters. The tomograms do show excess emission below the spot.  This behavior has been observed before in V801 Ara and in several similar systems (X1822-371, EXO 0748-676, Sco X-1, and V926 Sco). The lack of a hot spot could be due to a reduced accretion rate, and the enhanced emission below the spot where the accretion stream hits the disk implies that the stream is re-impacting the disk at a further point.

The tomograms constructed using H$\alpha$, H$\beta$, and He II $\lambda4686$ identify different regions of the accretion disk. H$\beta$ emission originates closer to the compact object and H$\alpha$ and He II originate from further towards the outer edge of the disk. This differentiates temperature variance in the disk for the first time.  The H$\alpha$ tomogram, created with our strongest, cleanest line, shows emission that appears to be centered on the center-of-mass of the system as has been observed for the similar systems X1822-371 and XTE J2123-058 as well as for systems containing black holes \citep{2008MNRAS.384.849N}.  This suggests the presence of an eccentric disk which could lead to disk precession.  Disk precession is known to occur in several neutron star LMXBs such as the X-ray pulsar Her X-1 \citep{1976ApJ...209..562G} and the Z-source Cyg X-2 \citep{1996ApJ...473L..45W}.  To confirm this for V801 Ara and for the other systems that show this behavior, considerably more observations taken at intervals that cover substantial portions of possible precession periods would be required.

\section{Acknowledgments}

We thankfully acknowledge the use of the MOLLY and DOPTOM software written by T. R. Marsh and the MODMAP software written by D. Steeghs.  This work is based on observations obtained with the Magellan/Baade telescope at Las Campanas Observatory.  K. Brauer was an REU intern at SAO while conducting this research.  The SAO REU program is funded in part by the National Science Foundation REU and Department of Defense ASSURE programs under NSF Grant no. 1262851, and by the Smithsonian Institution.







\newpage

\appendix

\renewcommand{\thetable}{A\arabic{table}}
\setcounter{table}{0}
\renewcommand{\thefigure}{A\arabic{figure}}
\setcounter{figure}{0}

\begin{figure*}
\centering
    \includegraphics[width=0.75\textwidth]{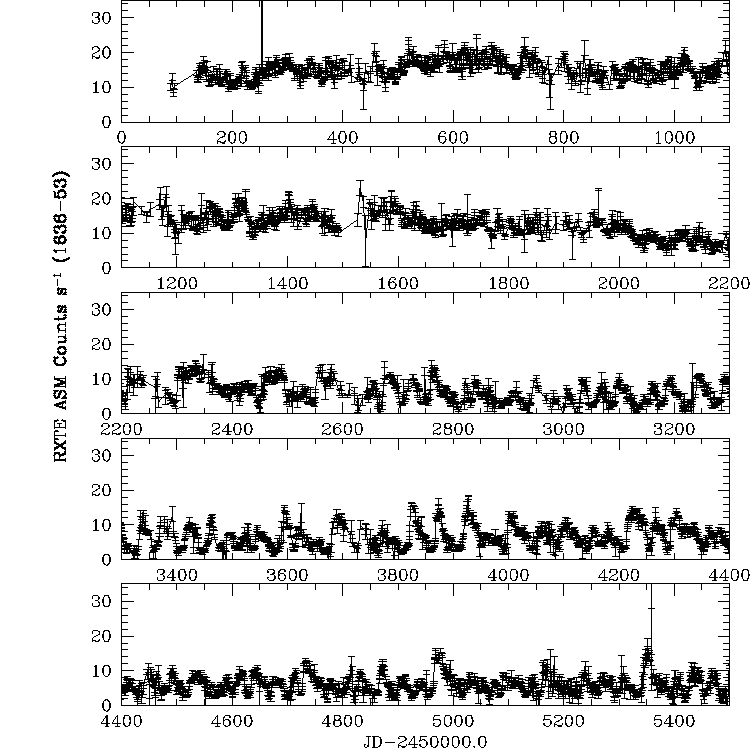}
    \caption{15 year lightcurve for V801 Ara using data from RXTE/ASM \citep{1996ApJ...469L..33L}.  The X-ray count rate, a measure of the accretion rate, varies from 2-20 RXTE counts with frequent excursions between 2-15. \label{fig:rxte}}
\end{figure*}

\begin{table}
\centering
    \caption{Observations of V801 Ara on 2014 May 30. Phases were computed using the ephemeris of \citet{2006MNRAS.373.1235C}. The error in our phase determination using this ephemeris is $\pm 0.02$ orbital cycles. \label{tab:obs}}
    \begin{tabular}{ccc}
   \hline
   UTC Time at  & Exposure & Phase at  \\
   Mid-Exposure (hr) & Time (s) & Mid-Exposure \\
    \hline
    2.6447222 & 400 & 0.474  \\
    2.8066667 & 600 & 0.517  \\
    2.9816667 & 600 & 0.563  \\
    3.1566667 & 400 & 0.609  \\
    3.2752778 & 600 & 0.641  \\
    3.4502778 & 600 & 0.687  \\
    3.6250000 & 600 & 0.733  \\
    3.8000000 & 600 & 0.779  \\
    3.9750000 & 600 & 0.825  \\
    4.1952778 & 600 & 0.883  \\
    4.3700000 & 600 & 0.929  \\
    4.5450000 & 600 & 0.975  \\
    4.7200000 & 600 & 1.022  \\
    4.8950000 & 600 & 1.068  \\
    5.0700000 & 600 & 1.114  \\
    5.2450000 & 600 & 1.160  \\
    5.4200000 & 600 & 1.206  \\
    5.6572222 & 600 & 1.269  \\
    5.8322222 & 600 & 1.315  \\
    6.0072222 & 600 & 1.361  \\
    6.1822222 & 600 & 1.407  \\
    6.3572222 & 600 & 1.453  \\
    6.5808333 & 600 & 1.512  \\
    6.7558333 & 600 & 1.558  \\
    6.9830556 & 600 & 1.618  \\
    7.1580556 & 600 & 1.664  \\
    7.3330556 & 600 & 1.710  \\
    7.5080556 & 600 & 1.757  \\
    7.6830556 & 600 & 1.803  \\
    7.9030556 & 600 & 1.861  \\
    8.0777778 & 600 & 1.907  \\
    8.2527778 & 600 & 1.953  \\
    8.4277778 & 600 & 1.999  \\
    8.6027778 & 600 & 2.045  \\
    8.8325000 & 600 & 2.106  \\
    9.0075000 & 600 & 2.152  \\
    9.1825000 & 600 & 2.198  \\
    9.3575000 & 600 & 2.244  \\
    9.5322222 & 300 & 2.290  \\
    \end{tabular}
\end{table}


\bsp    
\label{lastpage}
\end{document}